\documentclass[reviewcopy]{elsarticle}
\usepackage[reviewcopy]{adndt}
\usepackage{longtable}

\usepackage{amsmath}
\usepackage{color}
\usepackage{soul}
\usepackage{ulem}
\usepackage{amssymb}
\usepackage{graphicx}
\usepackage{pdflscape}
\usepackage{booktabs}
\usepackage{tabularx}
\biboptions{square,sort&compress}
\bibpunct[]{[}{]}{,}{n}{}{;}
\newcommand{\dprime}{{\prime\prime}}

\begin{document}

\begin{frontmatter}

\journal{Atomic Data and Nuclear Data Tables}


\title{Energy Levels, Transition Rates and Electron Impact Excitation Rates for B-like Kr XXXII}

\author[IMP]{Y. T. Li}
\author[IMP]{R. Si\corref{corl}}
\ead{rsi13@fudan.edu.cn}
\author[IMP]{J. Q. Li}
\author[IMP]{C. Y. Zhang}
\author[IMP]{K. Yao\corref{corl}}
\ead{keyao@fudan.edu.cn}
\author[HB]{K. Wang\corref{corl}}
\ead{wang$_{-}$kai10@fudan.edu.cn}
\author[KF]{M. F. Gu}
\author[IMP]{C. Y. Chen}
\cortext[corl]{Corresponding Author}

\address[IMP]{Shanghai EBIT Lab, Institute of Modern Physics, Department of Nuclear Science and Technology, Fudan University, Shanghai 200433, China}
\address[HB]{Hebei Key Lab of Optic-electronic Information and Materials, The College of Physics Science and Technology, Hebei University, Baoding 071002, China}
\address[KF]{Space Science Laboratory, University of California, Berkeley, CA 94720, USA}

\begin{abstract}
Energy levels and transition rates for electric-dipole, electric-quadrupole, electric-octupole, magnetic-dipole, and magnetic-quadrupole transitions among the levels arising from the $n\ \leq$ 5 configurations in B-like Kr \uppercase\expandafter{\romannumeral32} are calculated by using two state-of-the-art methods, namely, the multi-configuration Dirac-Hartree-Fock (MCDHF) approach and the second-order many-body perturbation theory (RMBPT). Our results are compared with several available experimental and other theoretical values. Electron-impact excitation (EIE) collision strengths are calculated via the independent process and isolated resonance approximation using distorted-wave (denoted by IPIRDW). Radiation damping effects on the resonance excitation contributions are included. Effective collision strengths are calculated as a function of electron temperature by assuming a Maxwellian electron velocity distribution. Spectral line intensities are modeled by using collision radiative model, and several line pairs pointed out might be useful for density diagnostics.
\end{abstract}

\end{frontmatter}




\newpage

\tableofcontents \vskip4pc


\section{Introduction}
It is well known that atomic spectroscopy of highly-charged ions is of particular interest for plasma control and plasma diagnostics on chemical composition, electron density and temperature, and electric and magnetic fields. Krypton, as a rare gas, has been widely used in diagnosing tokamak fusion plasmas~\cite{wyart1985identification, denne1989spectrum, jupen1990transitions}, since it does not pollute the vacuum vessel and is easily introduced into the plasmas. Furthermore, due to its almost ideal ionization balance and ease of injection, krypton has been the prime candidate for core Doppler spectroscopy on ITER~\cite{post1997progress,barnsley2004design,donne2007diagnostics}. Atomic data (including energy levels, transition rates, collision strengths, etc.) are required for many krypton ions. Recently, we reported accurate level energies, transition parameters and lifetimes for Kr \uppercase\expandafter{\romannumeral24}~\cite{zhang2018calculations}, Kr \uppercase\expandafter{\romannumeral25}~\cite{Si2015} and Kr XXVII -- Kr XXXI~\cite{Wang2016,zhang2018extended,wang2017extended,wang2017calculations,wang2014systematic}, using both the RMBPT approach and MCDHF method. This work reports our efforts for B-like Kr \uppercase\expandafter{\romannumeral32}.

The experimental studies on B-like Kr were very limited~\cite{denne1989spectrum,martin19902s,wyart1985identification,myrnas1994transitions,Podpaly2014,kukla2005extreme}, and only 11 $n=2$ levels have been compiled by the Atomic Spectra Database (ASD) of the National Institute of Standards and Technology (NIST)~\cite{kramida2015nist}. Most of the existing theoretical calculations were confined to the $n=2$ and $n=3$ configurations~\cite{safronova1996relativistic,Bogdanovich2007,Li_2010,Marques2012,Artemyev2013,naze2014isotope,verdebout2014hyperfine,Aggarwal2008,fontes2014relativistic,zhang1994relativistic}. Of particular note was the work presented by Aggarwal et al.~\cite{Aggarwal2008}, in which the Flexible Atomic Code (FAC)~\cite{gu2003indirect} and GRASP0 package~\cite{dyall1989grasp} have been adopted to obtain level energies, lifetimes, wavelengths and transition rates for transitions among the lowest 125 levels of the $n\leq3$ configurations. In addition, Aggarwal et al.~\cite{Aggarwal2009} also adopted the fully relativistic Dirac Atomic R-matrix Code (DARC) of Norrington and Grant to calculate collision strengths and effective collision strengths. Recently, Liang et al.~\cite{Liang2012} adopted Thomas-Femi-Dirac-Amaldi model potential from AUTOSTRUCTURE (AS)~\cite{badnell1986dielectronic} to calculate the energy levels, line strengths and lifetimes for the 204 $n\leq4$ energy levels of several B-like ions including Kr XXXII. Meanwhile, Liang et al.~\cite{Liang2012} performed intermediate-coupling frame transformation (ICFT) R-matrix calculation of electron-impact excitation (EIE) among the 204 states of Kr XXXII. However, the available theoretical energy levels and transition data~\cite{Aggarwal2008,Liang2012} were limited to the low-lying $n = 3, 4$ levels. The collisional data provided by Aggarwal et al.~\cite{Aggarwal2009} didn't include enough resonance contributions and ignored the radiative damping effect. The ICFT R-matrix method employed by Liang et al.~\cite{Liang2012} was not fully relativistic and also ignored the radiative damping effect. In our previous work~\cite{si2018energy}, we reported extensive and highly accurate energy levels, transition rates and electron impact excitation rates for B-like ions with $Z=24-30$, and discussed the importance of the radiative damping effect. The accuracy of energy levels was comparable to those performed by the relativistic Multi-Reference M${\o}$ller-Plesset Perturbation theory (MR-MP)~\cite{santana2018relativistic,san2018relativistic,santana2019relativistic}. In this work, we extend our previous research to $Z=36$, as a continuation of our recent efforts~\cite{si2018energy,Liu2019,wang2017extended,wang2018energy,wang2014systematic,wang2017calculations,si2017electron,si2016extended, li2015radiative,Li2015,Li2017} to provide self-consistent, large-scale and accurate atomic data.

In the current research, the RMBPT approach implemented within the FAC package~\cite{gu2008flexible} is adopted to calculate all the 513 energy levels that arise from the $2l^3$, $2l^23l^\prime$, $2l^24l^\prime$, and $2l^25l^\prime$ configurations of Kr \uppercase\expandafter{\romannumeral32}. E1, E2, E3, M1, and M2 transition rates among these levels are also reported. In order to assess the accuracy of the RMBPT results, we have also performed the MCDHF calculations using the GRASP2K package~\cite{jonsson2013new} for the lowest 300 levels arising from the $2l^3$, $2l^23l^\prime$, $2l^24l^\prime$ and $2s^25l$ configurations. In addition, the EIE effective collision strengths between the above 513 levels are calculated using the IPIRDW approximation over a wide temperature range from $5.12\times10^{5}$~K to $2.05\times10^{9}$~K. Based on the present atomic data, emission-line intensities are simulated using a collisional radiative model (CRM). Extensive comparisons are also made with the experimental and other theoretical values to assess the accuracy of our data.

\section{Calculations}
\subsection{\rm{RMBPT}}
In the RMBPT approach~\cite{Lindgren1974,Safronova1996}, implemented by Gu~\cite{gu2005wavelengths,gu2005energies,gu2006inner} in FAC, the Hamiltonian for an $n$-electron atom or ion is the no-pair Dirac-Coulomb-Breit (DCB) Hamiltonian ($H_{\rm{DCB}}$), which includes the frequency-independent Breit interaction. Separating the full Hilbert space into an orthogonal space $N$ and model space $M$ is the key to the RMBPT approach. By solving the effective Hamiltonian in the model space $M$, we can obtain the eigenvalues of the full Hamiltonian to the second order. This approach guarantees that electron correlation within the model space $M$ is exactly accounted for while the interactions between the model space $M$ and the orthogonal space $N$ are treated as a perturbation. In the present work, the model space $M$ contains all the configurations arising from  $2l^3$, $2l^23l^\prime$, $2l^24l^\prime$, and $2l^25l^\prime$. All the possible configurations generated by single and double (SD) excitations from the $M$ space are contained in the space $N$. Small corrections such as the leading QED corrections (electron self-energy and vacuum polarization) are also included.

\subsection{\rm{MCDHF}}
In the MCDHF method~\cite{Fischer2016} implemented in GRASP2K package~\cite{jonsson2013new}, the calculation starts with the Dirac-Coulomb Hamiltonian. The atomic state wavefunction (ASF) is expressed as an expansion over configuration state functions (CSFs) with the same parity ($P$) and angular momentum ($J$) by allowing single and double (SD) excitations from the reference configurations. Both the expansion coefficients and the radial parts of Dirac orbitals are optimized in the relativistic self-consistent field procedure. Breit and QED corrections are introduced in the relativistic configuration interaction (RCI) calculation~\cite{mackenzie1980program}, without changing the orbitals. In the present work, the reference configurations are same as those in the $M$ space of the RMBPT calculations, i.e.,  $2l^3$, $2l^23l^\prime$, $2l^24l^\prime$, and $2l^25l^\prime$ configurations. Valence-valence (VV) correlation and core-valence (CV) correlations are accounted for by allowing SD excitations from the above reference configurations to active sets with $n\leq8$, $l\leq7$.

\subsection{\rm{IPIRDW}}
In the IPIRDW approximation, contributions from direct excitation (DE) and resonance excitation (RE) to the total EIE effective collision strengths are obtained independently. Although the interference effect is ignored in the IPIRDW method, it has been shown in many cases that this method can give results in good agreement with those from the R-matrix for highly charged ions~\cite{gu2004electron,landi2006atomic,si2017electron,Chen2010}.

In the calculation of DE collision strengths, all the partial wave contributions up to $l=75$ are included explicitly, while higher $l$ contributions are included using the Coulomb-Bethe approximation~\cite{Burgess1970,burgess1974electron}. The DE collision strengths are calculated at six scattered electron energies, 1069.29 Ry times 0.001, 0.004, 0.02, 0.08, 0.2, and 0.6, using the relativistic distorted wave (RDW) approximation. We also compute the high-energy cross-sections at additional three-scattered energies of 1069.29 Ry times 8, 30, and 100 employing the relativistic plane-waves (RPW) approximation~\cite{Fontes2007}. Using the asymptotic behavior of the reduced collision strengths for different types of transitions, namely allowed and forbidden electric dipole transitions, we could extrapolate/interpolate DE collision strengths at higher energies (see for example Refs.~\cite{Chen2010,Wang2011,Li2015,si2017electron} for more details). DE effective collision strengths ($\Upsilon$) are obtained by integrating collision strengths over a Maxwellian distribution of electron velocities.

RE contributions through the relevant C-like doubly excited configurations $1s^{2}2l^{2}n^{\prime}l^{\prime}n^{\dprime}l^{\dprime}$ ($l\leq1,\ n^{\prime}\leq5,\ l^{\prime}\leq(n-1),\ n^{\dprime}\leq75,\ l^{\dprime}\leq8$) are included using the IPIRDW approximation. The radiative damping decays from C-like doubly excited states to resonant stabilizing (RS) states, as well as decays to low-lying autoionizing levels that may be followed by autoionization cascades (DAC) are taken into account. More details can be found in our previous studies~\cite{Shen2007,Chen2010,si2017electron}.

\section{Results and Discussion}
\subsection{Energy Levels}
The excitation energies for the lowest 513 states of the $2l^3$, $2l^23l^\prime$, $2l^24l^\prime$, and $2l^25l^\prime$ configurations from our RMBPT calculations, and the results for the lowest 300 states of the $2l^3$, $2l^23l^\prime$, $2l^24l^\prime$ and $2s^25l$ configurations from our MCDHF calculations are listed in Table~\ref{tab_energy}. The MCDHF and RMBPT wavefunctions are originally expressed in terms of the $jj$-coupled CSFs. In order to match the computed states against the NIST ASD and other calculations, the representations of CSFs are transformed from the $jj$-coupling to the $LSJ$-coupling using the program provided by Gaigalas et al.~\cite{gaigalas2004spectroscopic,gaigalas2017jj2lsj}. For each state numbered by a key, the $LSJ$-coupled and $jj$-coupled labels, as well as the $LSJ$-coupled and $jj$-coupled mixing coefficients for the lowest 513 levels from present RMBPT calculations are displayed in Table~\ref{tab_energy} (In the text, only the $LSJ$-coupled labels and mixing coefficients are listed. The $jj$-coupled labels and mixing coefficients are given in the supplements). From the mixing coefficients, we can see that many levels are strongly mixed, for which there are no unique $LSJ$ identifications, such as 270/271/272 and 286/289/292. The corresponding purity can be calculated by the mixing coefficient of $LSJ$-coupled and $jj$-coupled basis. The average $jj$ purity obtained in this work is found to be equal to 90\%, while the average $LS$ purity is found to be equal to 57\%. As expected, for such highly ionized atoms, due to the strong relativistic effects, the $jj$ coupling is much more suitable than the $LS$ coupling.

\begin{figure}[ht!]
 \centering
 \includegraphics[height=4.5in]{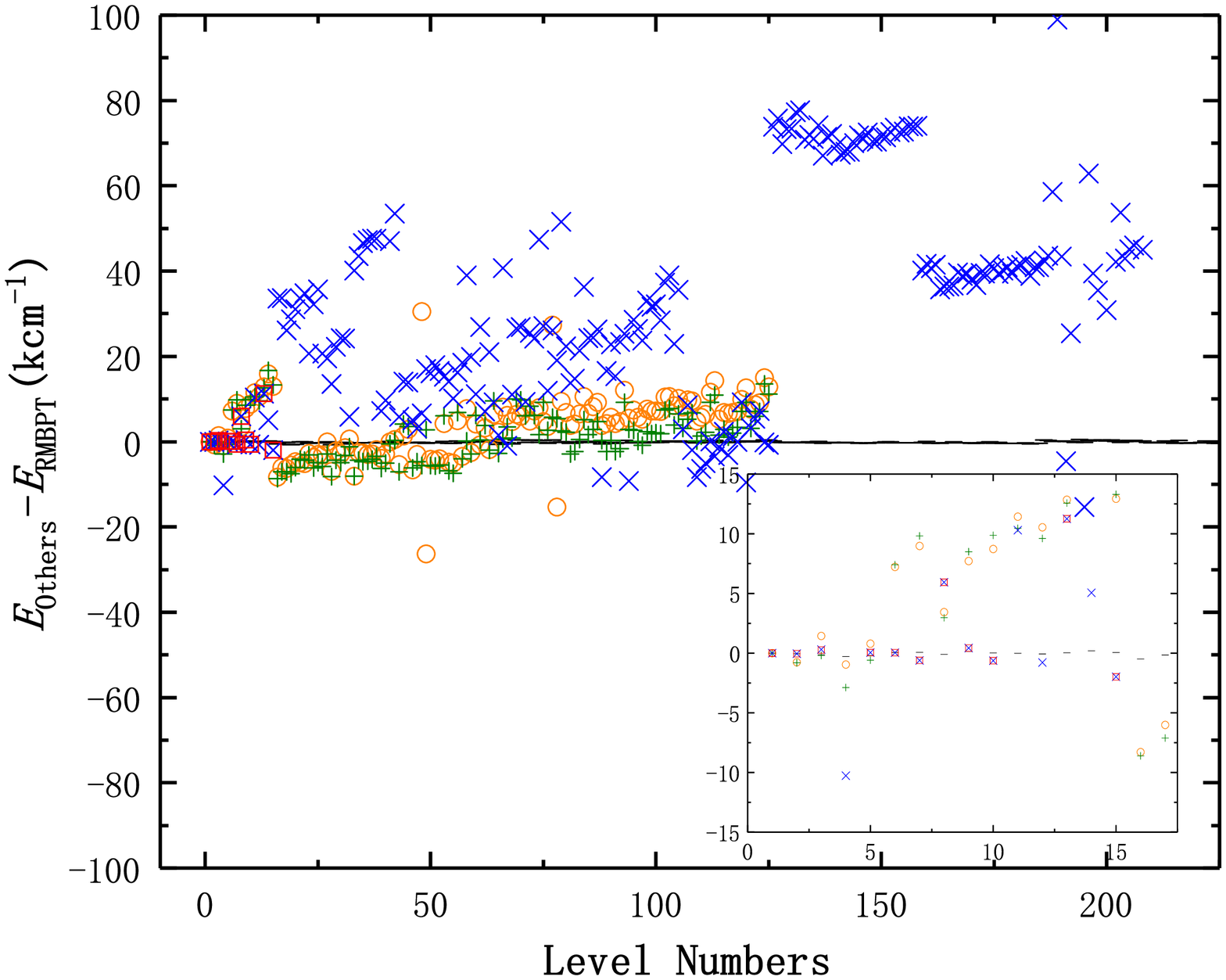}
 \caption{Comparisons of excitation energies from the present RMBPT calculations to those from the present MCDHF calculation (`` - ''), the compiled experimental values from the NIST ASD~\cite{kramida2015nist} (`` $\square$ ''), the calculations of Aggarwal et al.~\cite{Aggarwal2008} using the FAC code (`` $\circ$ '') and GRASP0 code (`` + ''), and the calculations of Liang et al.~\cite{Liang2012} using the AS code (`` $\times$ ''). The inset figure shows the differences for the $n = 2$ levels.}
 \label{fig_energy}
\end{figure}

In Fig.~\ref{fig_energy} and Table~\ref{tab_compare_energy}, the present RMBPT and MCDHF level energies are compared with the GRASP0 and FAC results~\cite{Aggarwal2008} and the AS ones~\cite{Liang2012}. It can be seen that the present two sets of energies for the lowest 300 levels are in excellent agreement by within 0.01\%, with only three exceptions up to 0.03\% for the $n=2$ levels ($2s(^{2}S)2p^{2}\ ^{4}P_{1/2}$, $ 2s(^{2}S)2p^{2}\ ^{4}P_{5/2}$ and $ 2s(^{2}S)2p^{2}\ ^{4}P_{3/2}$). The mean of the relative differences, with its standard deviation, is 8ppm $\pm$ 30ppm. The present RMBPT and MCDHF results support each other and certify the reliability of our calculations. The NIST energy (also listed in Table~\ref{tab_compare_energy}) for $2s(^{2}S)2p^{2}\ ^{2}S_{1/2}$ ($E_{\rm{NIST}} = 1502900$ cm$^{-1}$) is in much better agreement with our 7th level ($2s(^{2}S)2p^{2}\ ^{2}P_{1/2}$, $E_{\rm{RMBPT}} = 1503506$ cm$^{-1}$) than the 9th level ($2s(^{2}S)2p^{2}\ ^{2}S_{1/2}$, $E_{\rm{RMBPT}} = 2029008$ cm$^{-1}$), while $2s(^{2}S)2p^{2}\ ^{2}S_{1/2}$ ($E_{\rm{NIST}} = 2029440$ cm$^{-1}$) is in much better agreement with our 9th level than the 7th level, thus we interchanged the two NIST level identifiers. In addition, 3 out of the 11 NIST~\cite{kramida2015nist} values differ from the present calculations by over 1000 cm$^{-1}$. The largest difference occurs in $2p^3\  ^2D_{5/2}$ ($E_{\rm{RMBPT}} = 2732041$ cm$^{-1}$, $E_{\rm{NIST}} = 2743300$ cm$^{-1}$) and $2s2p^2\  ^2D_{5/2}$ ($E_{\rm{RMBPT}} = 1670690$ cm$^{-1}$, $E_{\rm{NIST}} = 1676630$ cm$^{-1}$), the deviations are due to the differences in the transition wavelengths, and the details will be discussed in the next section.
As for the energy levels from other theoretical calculations, only 8\% of FAC~\cite{Aggarwal2008} results, 17\% of GRASP0~\cite{Aggarwal2008} and 5\% of AS~\cite{Liang2012} results agree with our results by within 0.01\%, the maximum deviation is up to 1\%.

\subsection{Wavelengths}
Our RMBPT and MCDHF wavelengths for E1, E2, E3, M1 and M2 transitions with branching ratios over 0.1\% among the levels in Table~\ref{tab_energy} are listed in Table~\ref{tab_tr}. All the wavelengths from the present MCDHF and RMBPT calculations agree within 0.1\%, and 96\% of them agree within 0.01\%. The mean relative difference is -0.2ppm $\pm$ 90ppm.  This is highly satisfactory.

\begin{figure}[ht!]
\centering
 \includegraphics[height=4.5in]{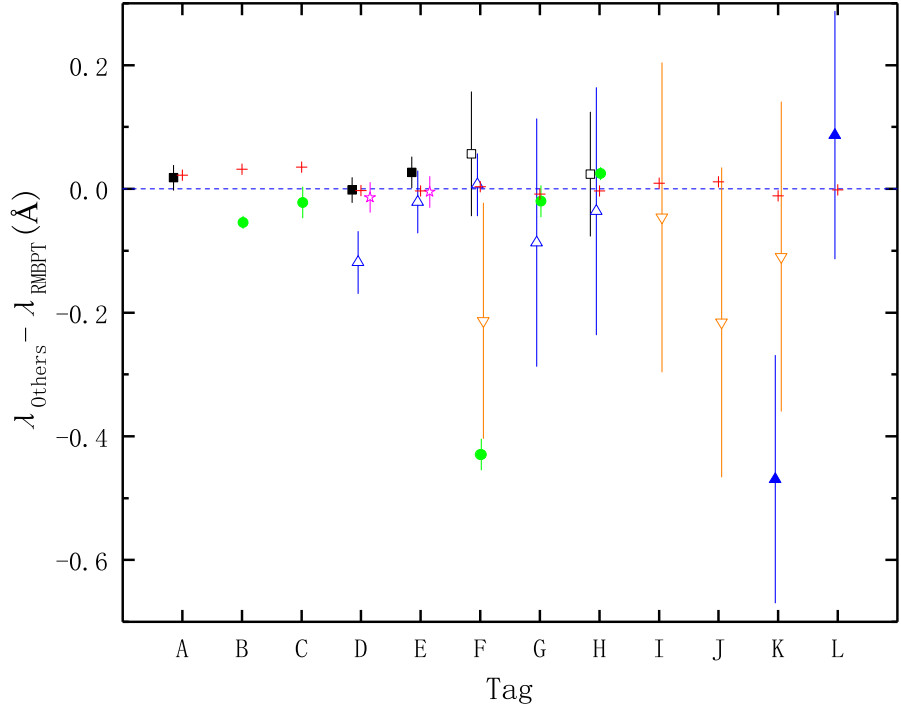}
\caption{Comparisons of wavelengths from the present RMBPT calculations to those from the present MCDHF calculations (plus), the observed wavelengths from Denne et al.~\cite{denne1989spectrum} (square), Martin et al.~\cite{martin19902s} (up triangle), Myrn{\"a}s et al.~\cite{myrnas1994transitions} (circle), Podpaly~\cite{Podpaly2014} (star) and Kukla et al.~\cite{kukla2005extreme} (down triangle), respectively. The experimental values compiled by the NIST ASD~\cite{kramida2015nist} are marked as solid symbols. The tags for transitions are listed in Table~\ref{tab_compare_wavelengths}. The horizontal lines indicate differences of 0 ${\rm{\AA}}$.}
 \label{fig_wavelengths}
\end{figure}

Denne et al.~\cite{denne1989spectrum}, Martin et al.~\cite{martin19902s}, Myrn{\"a}s et al.~\cite{myrnas1994transitions}, Podpaly et al.~\cite{Podpaly2014} and Kukla et al.~\cite{kukla2005extreme} have measured some wavelengths for Kr XXXII, and all of them are for $n=2\rightarrow 2$ transitions.
In Table~\ref{tab_compare_wavelengths} and Fig.~\ref{fig_wavelengths}, the present RMBPT and MCDHF wavelengths are compared with these observed values. For convenience, we label the transitions as A, B, C, etc. 
We can see that for a same transition, the observed wavelengths from different experiments can differ by up to 0.4\% (transition K), some of them can not coincide within each other's error bars (transition D), and accurate theoretical values are needed to distinguish them.
Our calculated wavelengths generally agree well with the experimental values from Ref.~\cite{denne1989spectrum}, and basically within the error bar of the experimental values from Ref.~\cite{kukla2005extreme}, although the latter observations are with relatively big uncertainties.
The NIST ASD compiled some of the experimental wavelengths, and most of the compilations agree well with our calculations. The two exceptions are the transition F [$2s(^{2}S)2p^{2}\ ^{2}D_{5/2} \rightarrow 2s^{2}2p \ ^{2}P_{3/2}$] and the transition K [$2p^{3}\ ^{2}D_{5/2} \rightarrow 2s(^{2}S)2p^{2}\ ^{2}D_{5/2}$]. The deviations in these two transition wavelengths result in the large energy deviations in these two upper levels, as mentioned in section~3.1. For transition F with $\lambda_{\rm{NIST}} = 84.454 \pm 0.025\ {\rm{\AA}}$, $\lambda_{\rm{RMBPT}} = 84.8834\ {\rm{\AA}}$, and $\lambda_{\rm{MCDHF}} = 84.8868\ {\rm{\AA}}$, our calculations are far out of the NIST complication's uncertainty, but agree well with the observed values from Ref.~\cite{martin19902s} ($84.89 \pm 0.05\ {\rm{\AA}}$) and Ref.~\cite{denne1989spectrum} ($84.94 \pm 0.10\ {\rm{\AA}}$).
For transition K, our calculated wavelengths ($\lambda_{\rm{RMBPT}} = 94.2195\ {\rm{\AA}}$, $\lambda_{\rm{MCDHF}} = 94.2082\ {\rm{\AA}}$) agree within the observed value from Ref.~\cite{kukla2005extreme} ($\lambda = 94.11\pm0.25\ {\rm{\AA}}$), but out of the NIST complication's uncertainty ($\lambda_{\rm{NIST}} = 93.75\pm0.2\ {\rm{\AA}}$).
We can see that without accurate theoretical calculations, it is not easy to say which experimental value is more reliable.

\subsection{Transition Rates and Lifetimes}
Table~\ref{tab_tr} lists our RMBPT and MCDHF transition rates ($A$, in s$^{-1}$) for E1, E2, E3, M1 and M2 transitions with branching ratio over 0.1\% between the levels of Table~\ref{tab_energy}. For different types of transitions, line strength ($S$, in a.u.) and oscillator strength ($f$, dimensionless) can be obtained by radiation rate $A$ as follows (these values are included in the supplements):

For the E1 transition:
\begin{equation}
A_{ji}=\frac{2.0261\times10^{18}}{\omega_{j}\lambda^{3}_{ji}}S, f=\frac{303.75}{\lambda_{ji}\omega_{i}}S \; ,
\end{equation}

for the M1 transition:
\begin{equation}
A_{ji}=\frac{2.6974\times10^{13}}{\omega_{j}\lambda^{3}_{ji}}S, f=\frac{4.044\times10^{-3}}{\lambda_{ji}\omega_{i}}S \; ,
\end{equation}

for the E2 transition:
\begin{equation}
A_{ji}=\frac{1.1199\times10^{18}}{\omega_{j}\lambda^{5}_{ji}}S, f=\frac{167.89}{\lambda_{ji}\omega_{i}}S \; ,
\end{equation}

for the M2 transition:
\begin{equation}
A_{ji}=\frac{1.4910\times10^{13}}{\omega_{j}\lambda^{5}_{ji}}S, f=\frac{2.236\times10^{-3}}{\lambda_{ji}\omega_{i}}S \; ,
\end{equation}

and for the E3 transition:
\begin{equation}
A_{ji}=\frac{3.1444\times10^{17}}{\omega_{j}\lambda^{7}_{ji}}S, f=\frac{47.140}{\lambda_{ji}\omega_{i}}S \; ,
\end{equation}
where  $\omega_{i}$ and $\omega_{j}$ are the statistical weights of the lower ($i$) and upper ($j$) levels, respectively.

For electric multipole transitions, line strength ($S$) has three standard gauges of expressions, i.e., length, velocity, and acceleration. In the framework of relativity, the length and velocity gauges of electrical multipole transitions are given by the Babushkin and Coulomb metrics, respectively~\cite{1962OptSp..13...77B}. If the wave functions are accurate enough, the line strengths calculated by these three gauges should be the same. The degree of conformity of the calculation results under different gauges can be used as a way to check the accuracy of wave fuctions~\cite{fischer2009evaluating}. However, it should be mentioned that this type of inspection is necessary but not sufficient. In general, transition data for electric multipoles calculated in the Babushkin gauge are preferable over those calculated in the Coulomb gauge. In this work, the RMBPT line strengths are calculated under the length gauge, while the MCDHF results are calculated under both the length and velocity gauges.

\begin{figure}[ht!]
\center
\includegraphics[height=4.5in]{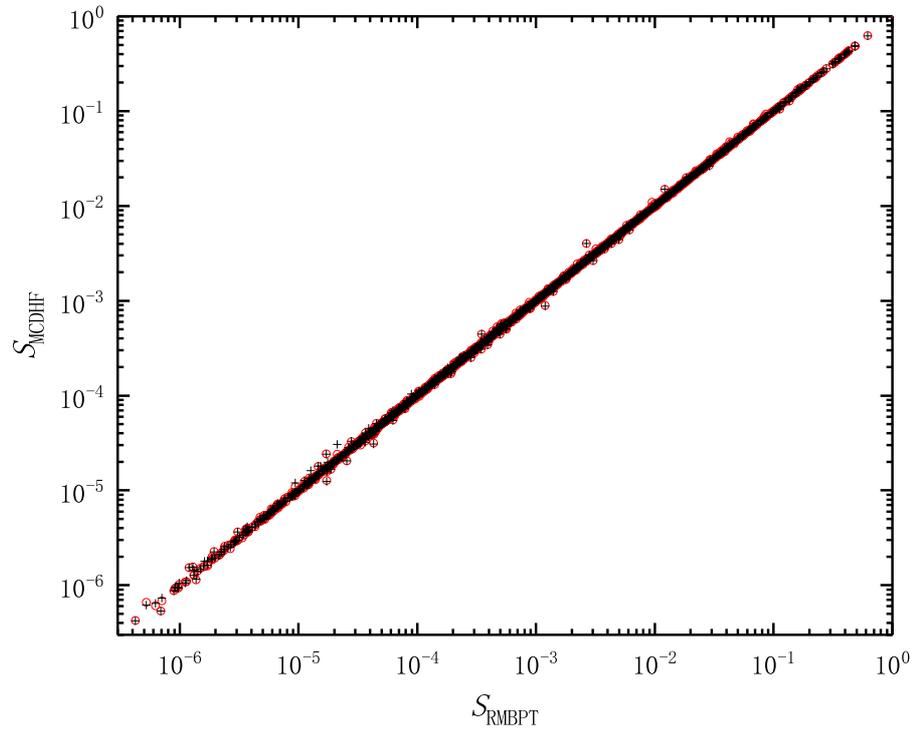}
\caption{Comparisons of line strengths $S$ for transitions from the present RMBPT calculations with those from present MCDHF calculations in the length gauge (`` + '') and velocity gauge (`` $\circ$ ''), respectively.}
 \label{fig_S}
\end{figure}

In Fig.~\ref{fig_S}, we compare the present RMBPT line strengths with the MCDHF values in both the length and velocity gauges.
We can see that more than 98\% of the present MCDHF results agree with the present RMBPT values by within 10\%, the length gauge and velocity gauge MCDHF values are also in very good agreement, which gives us confidence in our calculations.
Only 16 out of the 2,788 MCDHF values deviate from the present RMBPT values by more than 20\%. They are usually very weak, and the BRs are on the order of a few thousandths. Many of them are two-electron-one-photon transitions, such as 118$\rightarrow$1 [$ 2p^{2}(^{3}P)4s\ ^{4}P_{  1/2} \rightarrow 2s^{2}2p\ ^{2}P_{1/2}$] and 213$\rightarrow$2 [$ 2p^{2}(^{3}P)4s\ ^{2}P_{1/2} \rightarrow 2s^{2}2p\ ^{2}P_{3/2}$]. Only 2 transitions differ by more than one order of magnitude, and they are 287$\rightarrow$6 [$ 2s^{2}5p\ ^{2}P_{ 1/2} \rightarrow 2s(^{2}S)2p^{2}\ ^{2}D_{3/2}$] and 288$\rightarrow$8 [$ 2s^{2}5p\ ^{2}P_{  3/2}\rightarrow 2s(^{2}S)2p^{2}\ ^{2}D_{  5/2}$]. For them the corresponding MCDHF length and velocity gauges of line strengths also differ by two orders of magnitude.
These transitions are highly sensitive to electron correlation effects and calculation models, a slight difference in the calculations will lead to a relatively large difference in the computed $A$ values.

On the other hand, there are quite a lot of transition rates calculated by Aggarwal et al.~\cite{Aggarwal2008} differ from our RMBPT results by over 20\% even for the strong transitions listed in Table~\ref{tab_tr}. The large differences may be due to the fact that very limited electron correlations were included in their calculations.

\begin{figure}[ht!]
\center
\includegraphics[height=4.5in]{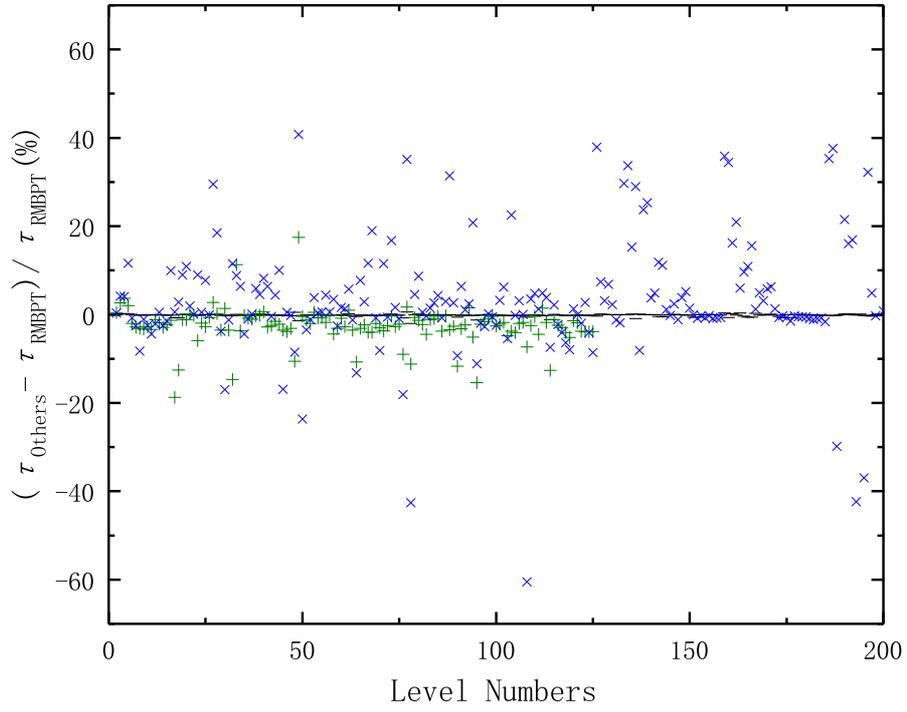}
\caption{Comparisons of lifetimes obtained from the present RMBPT calculations with those from the MCDHF calculations (`` - ''), and the calculations of Aggarwal et al.~\cite{Aggarwal2008} (`` + '') and Liang et al.~\cite{Liang2012} (`` $\times$ ''), respectively.}
 \label{fig_lifetimes}
\end{figure}

The radiative lifetime $\tau_{j}$ of a level $j$ is calculated from the radiative transition rates $A_{ji}$ using the relation
\begin{equation}
\tau_j=(\sum_i A_{ji})^{-1} \; ,
\end{equation}
In Table~\ref{tab_energy}, we list the present lifetimes from the RMBPT and MCDHF calculations. Since the MCDHF lifetimes in length and velocity gauges are in excellent agreement, only the ones in length gauge are listed in Table~\ref{tab_energy}. The relative deviations of the MCDHF, GRASP0~\cite{Aggarwal2008} and AS~\cite{Liang2012} lifetimes from the present RMBPT values are shown in Fig.~\ref{fig_lifetimes}. We can see that the two sets of lifetimes obtained from the present RMBPT and MCDHF calculations are in good agreement, and more than 96\% of them agree by within 1\% and all of them agree within 4\%. The mean of the relative differences, with its standard deviation, is -0.2\% $\pm$ 0.4\%. Some strongly mixing levels have a lifetimes difference of more than 1\%. For example, the levels 48/49 [$2s2p(^{3}P)3p\ ^{2}S_{  1/2} / 2p^{2}(^{3}P)3s\ ^{4}P_{  1/2}$] and 77/78 [$ 2s2p(^{1}P)3d\ ^{2}P_{  3/2} / 2p^{2}(^{3}P)3p\ ^{2}D_{  3/2}$].
As for the lifetimes from GRASP0 calculations~\cite{Aggarwal2008}, 13\% of results have a deviation from the present RMBPT ones beyond 5\%, with the maximum difference of 20\%. The lifetimes from AS calculations~\cite{Liang2012} show more obvious deviations from our RMBPT results, with only 60\% of the lifetimes agreeing with our RMBPT values within 5\%.

It can be seen from the radiation branch ratios (BRs) given in Table~\ref{tab_tr} that the lifetimes are mostly dominated by E1 transitions. The lifetime for the first excited-level $2s^{2}2p\ ^{2}P_{3/2}$ is determined by the M1 transition to the groundstate $2s^{2}2p\ ^{2}P_{1/2}$. The contributions of E2 transitions are also very important for many levels, for example, the BRs of E2 transitions from the $2s2p(^{3}P)3p\ ^{4}D_{7/2}$, $2p^{2}(^{3}P)3d\ ^{4}F_{9/2}$, $2p^{2}(^{1}D)3d\ ^{2}G_{9/2}$ and $2p^{2}(^{1}D)3d\ ^{2}F_{7/2}$ levels are as high as 86\%, 77\%, 68\% and 56\%, respectively. For $2s2p(^{3}P)3d\ ^{4}F_{9/2}$, the BR of M2 transition to $2s(^{2}S)2p^{2}\ ^{2}D_{5/2}$ exceeds 20\%, the BRs of E3 transition to $2s(^{2}S)2p^{2}\ ^{4}P_{3/2}$, $2s(^{2}S)2p^{2}\ ^{4}P_{5/2}$, $2s(^{2}S)2p^{2}\ ^{2}D_{5/2}$  are also greater than 0.1\%, and the sum of them is more than 3\%. The M3 transitions are very weak, and the largest M3 transition BR is only 0.02\% (for $2s2p(^{3}P)3d\ ^{4}F_{ 9/2}\rightarrow2s^{2}2p\ ^{2}P_{3/2}$), hence the M3 transitions are discarded in Table~\ref{tab_tr}.

\subsection{Effective Collision Strengths and Line Emissions}
In Table~\ref{tab_collisionstrengths}, we present our results of electron-impact excitation effective collision strengths between all the 513 levels of B-like Kr over a wide temperature range from $5.12 \times 10^{5}$ K to $2.05\times10^{9}$ K. The statistical weights 2$J$ + 1 of the upper and lower levels, the energy differences $\Delta E$, and total transition rates $A$ are also listed. Thus, this table is self-contained for a spectral calculation using collisional radiative model.

\begin{figure}[ht!]
\center
\includegraphics[height=4.5in]{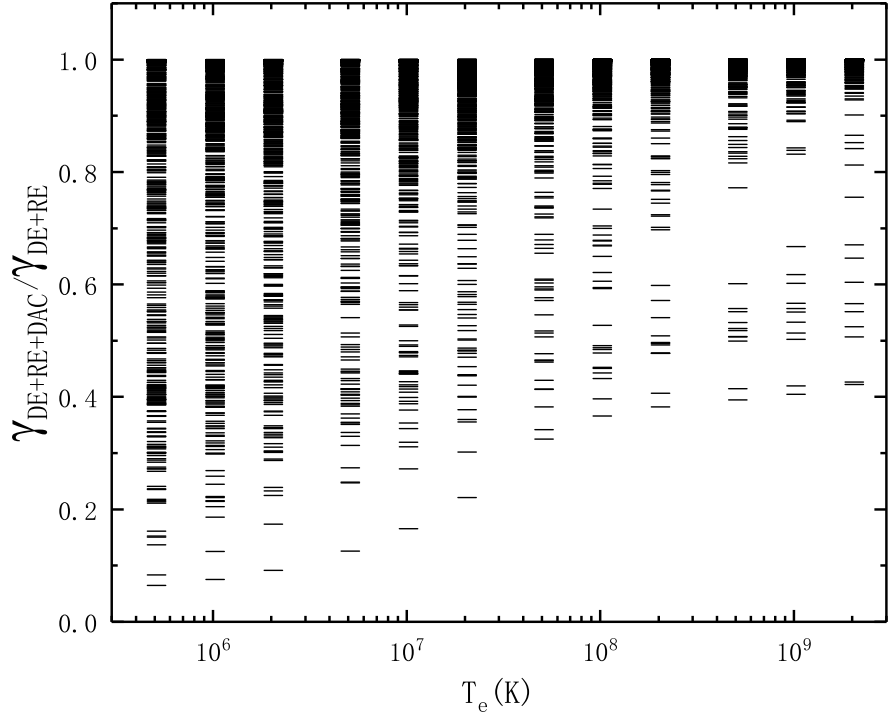}
\caption{The effect of radiative damping on effective collision strengths of the transitions initially from the ground state in Kr XXXII as a function of electron temperature.}
 \label{fig_damping}
\end{figure}

The radiative decay of the autoionizing state is a significant competitive process of autoionization. Using different approaches to include the radiative decay processes, the calculated RE collision strengths and the final collision strengths will differ. In the present work, the RS and DAC radiative processes are both taken into account. In Fig.~\ref{fig_damping}, we show the effect of radiative damping on the effective collision strengths of excitations for all the transitions from the ground state as functions of electron temperature. At the low-temperature end, effective collision strengths for over 70\% of the transitions are reduced by more than 10\%, although the effect decreases with increasing temperature, many effective collision strengths are still reduced by over 50\%, such as, for transitions 1 $\rightarrow$ 167, 181. It can be seen that in the middle and low-temperature areas, it is very necessary to consider radiative decays.

\begin{figure}[ht!]
\center
\includegraphics[height=4.5in]{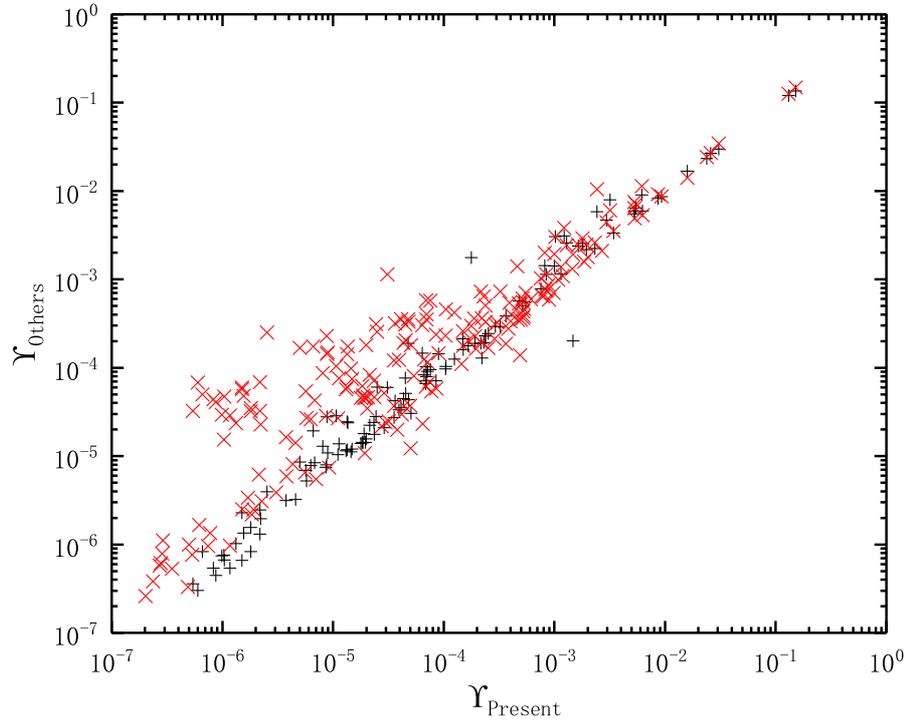}
\caption{Comparisons of effective collision strengths from the present calculations with those from Aggarwal et al.~\cite{Aggarwal2008} (``+'') and Liang et al.~\cite{Liang2012} (``$\times$'') for excitations originating from the ground state in Kr XXXII at $T_{e} = 1.02 \times 10^{6}  $ K, respectively.}
 \label{fig_compare_collisionstrengths}
\end{figure}

The present effective collision strengths results for the transitions among the 513 levels arising from the $2l^3$, $2l^23l^\prime$, $2l^24l^\prime$, and $2l^25l^\prime$ configurations differ significantly from the results reported in~\cite{Aggarwal2009} and those reported in~\cite{Liang2012}. As shown in Fig.~\ref{fig_compare_collisionstrengths}, we compare our computed effective collision strengths at $T_e$ = $1.02 \times 10^{6}$ K with those from~\cite{Aggarwal2009,Liang2012} for the excitations from the ground state. The results from~\cite{Aggarwal2009,Liang2012} are generally higher than the present values, and only 65\% and 34\% of effective collision strengths from Aggarwal et al.~\cite{Aggarwal2009} and Liang et al.~\cite{Liang2012} agree with our results to within 30\%. We should mention that the method used by Liang et al.~\cite{Liang2012} was nonrelativistic, and both of them~\cite{Aggarwal2009,Liang2012} ignored the radiative damping effect. Therefore, we believe our relativistic results including radiative damping effect are more reliable.

\subsection{Collisional Radiative Model Simulation}
By including spontaneous radiative transitions, collisional excitation and de-excitation processes between the present 513 levels in the CRM, and using the statistical equilibrium code of Dufton~\cite{dufton1977program}, we estimate the relative level populations and emission-line intensities. In Table~\ref{tab_CRM}, we present a summary list of the lines for which the relative intensities to the strongest line are larger than 0.1\%, at the electron temperature of 3.5 keV which is considered as the temperature of the largest abundance for B-like krypton~\cite{Mattioli2006} in the electron intensity range from $10^{12}$ to $10^{18}$ cm$^{-3}$.

\begin{figure}[ht!]
\center
\includegraphics[height=4.5in]{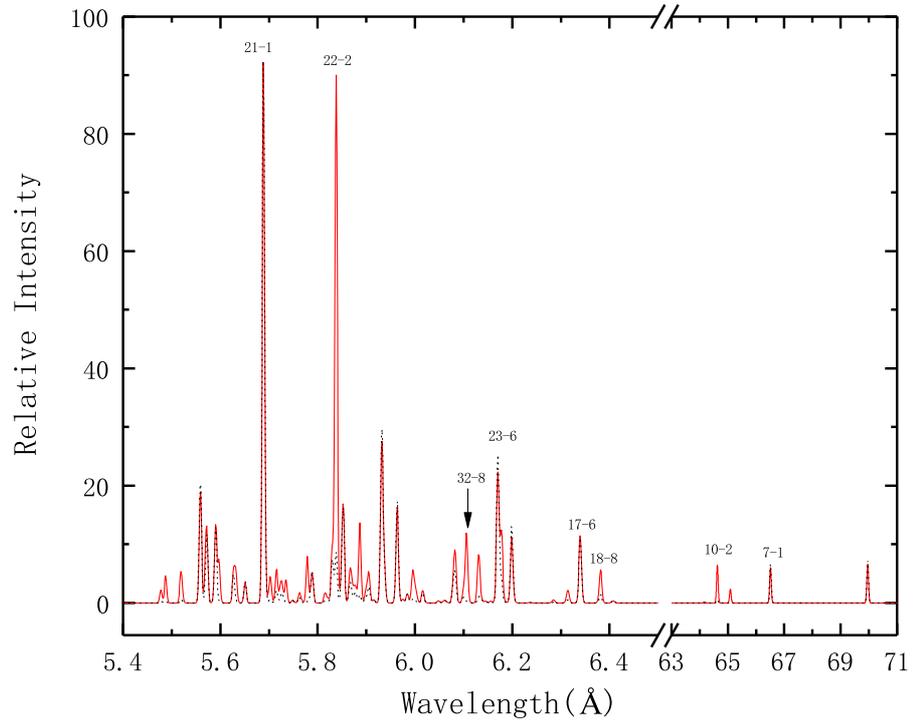}
\caption{Simulated spectra at a resolution of 1/1000 for transition wavelengths between 5.4 -- 6.5 ${\rm{\AA}}$ and 63 -- 71 ${\rm{\AA}}$ at the electron densities of $10^{13}\ \rm{cm}^{-3}$ (dashed line) and $10^{18}\ \rm{cm}^{-3}$ (solid line).}
 \label{fig_simulatespectra}
\end{figure}

\begin{figure}[ht!]
\includegraphics[width=10cm]{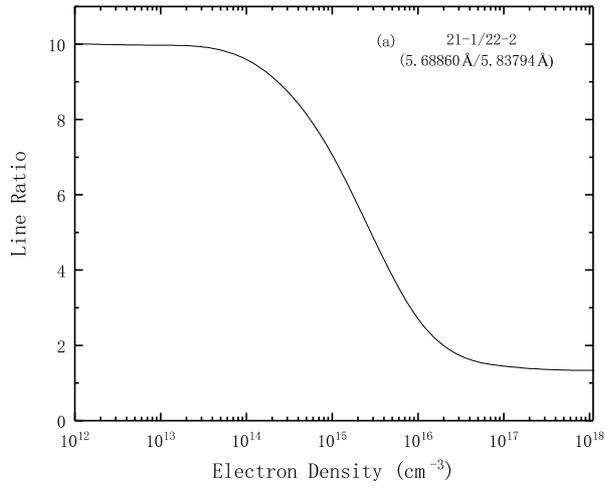}
\includegraphics[width=10cm]{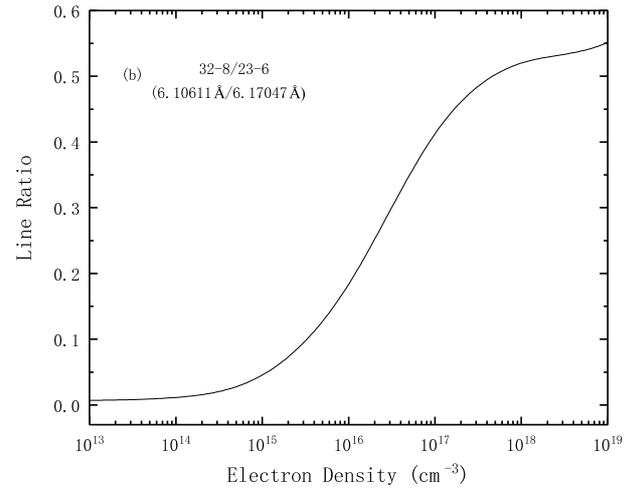}
\includegraphics[width=10cm]{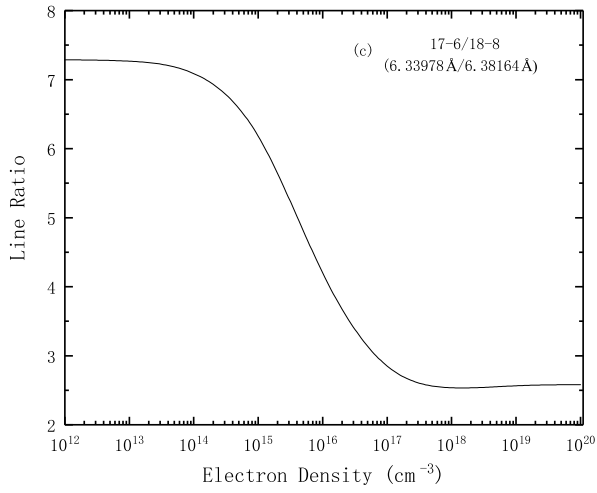}
\includegraphics[width=10cm]{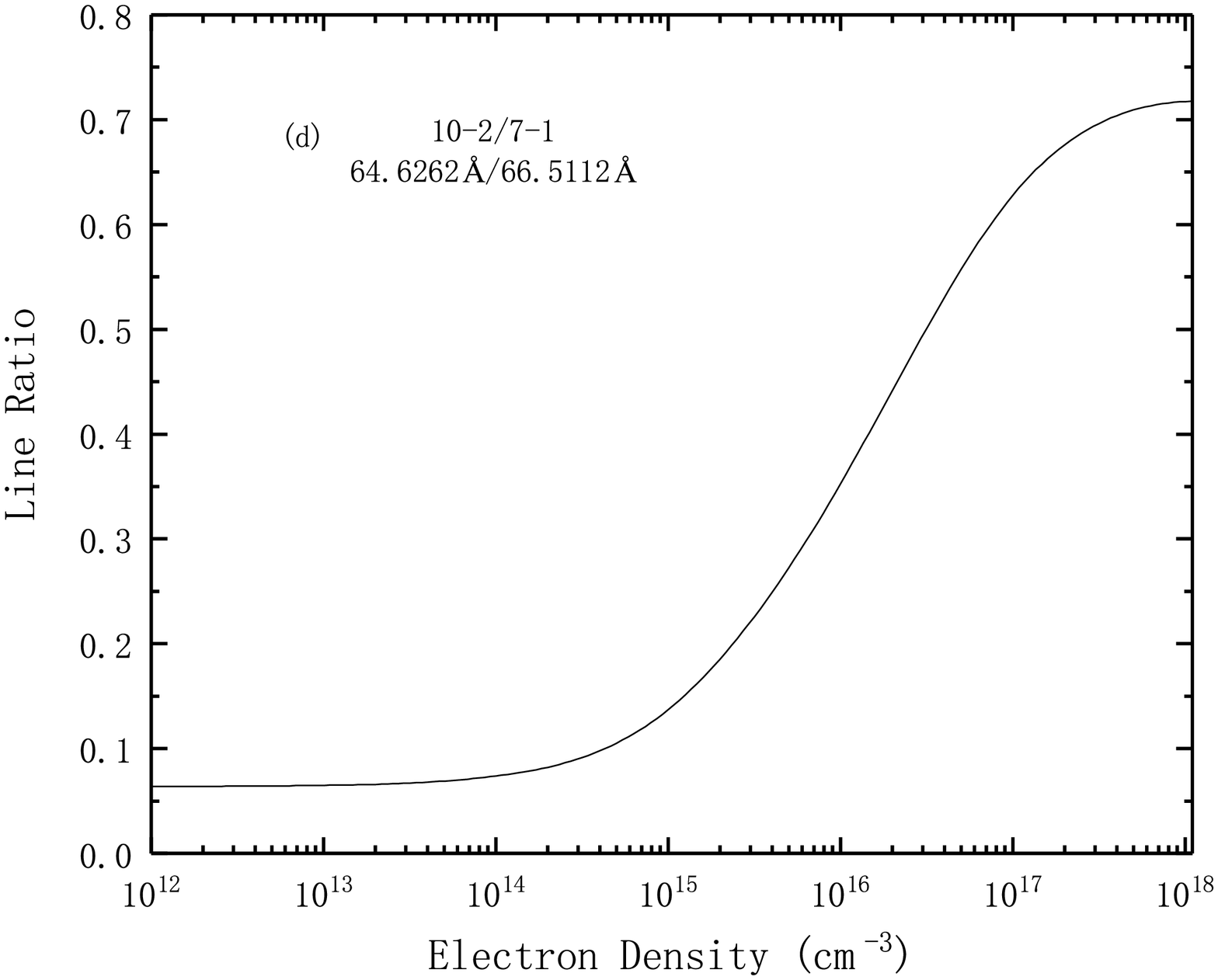}
\caption{Predicted intensity ratios for the selected line pairs.}
 \label{fig_predicted}
\end{figure}

In Fig.~\ref{fig_simulatespectra}, we show the simulated spectrum in the wavelength range 5.4 -- 6.5 ${\rm{\AA}}$ and 63 -- 71 ${\rm{\AA}}$. The solid and dashed lines represent the simulated relative intensity respect to the strongest line at the electron densities of $10^{13}\ \rm{cm}^{-3}$ and $10^{18}\ \rm{cm}^{-3}$, respectively. Some of the line pairs in Fig.~\ref{fig_simulatespectra} are marked. There are highly density-sensitive line pairs in the X-ray bands, which may be useful for plasma density diagnostics. They are for example $2s^{2}3d\ ^{2}D_{3/2} \rightarrow 2s^{2}2p\ ^{2}P_{1/2}$ (21 -- 1) and $2s^{2}3d\ ^{2}D_{5/2} \rightarrow 2s^{2}2p\ ^{2}P_{3/2}$ (22 -- 2), $2s2p(^{3}P)3s\ ^{2}P_{3/2} \rightarrow 2s(^{2}S)2p^{2}\ ^{2}D_{5/2}$ (32 -- 8) and $2s2p(^{3}P)3s\ ^{2}P_{1/2} \rightarrow 2s(^{2}S)2p^{2}\ ^{2}D_{3/2}$ (23 -- 6), $2s^{2}3p\ ^{2}P_{1/2} \rightarrow 2s(^{2}S)2p^{2}\ ^{2}D_{3/2}$ (17 -- 6) and $2s^{2}3p\ ^{2}P_{3/2} \rightarrow 2s(^{2}S)2p^{2}\ ^{2}D_{5/2}$ (18 -- 8), $2s(^{2}S)2p^{2}\ ^{2}P_{3/2} \rightarrow 2s^{2}2p\ ^{2}P_{3/2}$ (10 -- 2) and $2s(^{2}S)2p^{2}\ ^{2}P_{1/2} \rightarrow 2s^{2}2p\ ^{2}P_{1/2}$ (7 -- 1), characterized by density-sensitive line intensity ratios (see Fig.~\ref{fig_predicted}(a),(b), (c), and (d)).

\section{Conclusion}
We present energy levels as well as radiative rates for E1, E2, E3, M1 and M2 transitions among the 513 levels and 300 levels belonging to B-like Kr XXXII, using the RMBPT method and MCDHF approach, respectively. The two sets of results are in excellent agreement. Comparisons are also made with other theoretical results and experimental results. We also report electron impart excitation effective collision strengths calculated via the IPIRDW approximation over a wide electron temperature range from 5.12 $\times$ 10$^{5}$ K to 2.05 $\times$ 10$^{9}$ K. Some line pairs, which could be useful for plasma diagnostics are proposed. Our present results show an improvement in both accuracy and comprehensiveness over previous calculations, which can be used in line identifications, plasma modeling, and diagnostics for various laboratory plasmas.

\section*{Acknowledgments}
We acknowledge the support from the National Key Research and Development Program of China under Grant No.2017YFA0402300, and the National Natural Science Foundation of China (Grant No. 11703004 and No. 11674066) the Natural Science Foundation of HeBei Province, China (A2017201165). R. S. and K. W. express their gratitude for the support from the visiting researcher program at the Fudan University.



\bibliographystyle{./elsarticle-num}
\bibliography{B-Kr}

\clearpage
\section*{Figures}

\clearpage

\clearpage

\clearpage

\clearpage

\clearpage

\clearpage

\clearpage

\clearpage

\newpage
\setlength{\tabcolsep}{0.5\tabcolsep}
\renewcommand{\arraystretch}{0.7}
\footnotesize 
\begin{longtable}{@{\extracolsep\fill}llcrrrrrrrl@{}}
\caption{Comparisons between the energies (in cm$^{-1}$) relative to the ground state in Kr XXXII. $a$ - the NIST values~\cite{kramida2015nist}; $b$ - the present results; $c$ - Aggarwal et al.~\cite{Aggarwal2008}; $d$ - Liang et al.~\cite{Liang2012}. \vskip0.5pc
\label{tab_compare_energy}}
Key  & Conf. & Term &  NIST$^{a}$  & RMBPT$^{b}$ &  MCDHF$^{b}$   &  FAC$^{c}$  & GRASP0$^{c}$ & AS$^{d}$     \\
\hline\\
\endfirsthead\\
\caption[]{($continued$)}
Key  & Conf. & Term &  NIST$^{a}$  & RMBPT$^{b}$ &  MCDHF$^{b}$   &  FAC$^{c}$  & GRASP0$^{c}$ & AS$^{d}$     \\
\hline\\
\endhead
1	  &	$	2s^{2}2p	      $	&	$	^{2}P_{1/2}	$	&	0	      &	0	&	0	&	0	&	0	&	0	\\
2	  &	$	2s^{2}2p	      $	&	$	^{2}P_{3/2}	$	&	492560	&	492603	&	492551	&	491841	&	491789	&	492560	\\
3	  &	$	2s(^{2}S)2p^{2}	$	&	$	^{4}P_{1/2}	$	&	698000	&	697741	&	697582	&	699184	&	697541	&	698038	\\
4	  &	$	2s(^{2}S)2p^{2}	$	&	$	^{4}P_{3/2}	$	&		      &	1013778	&	1013493	&	1012827	&	1010894	&	1003503	\\
5	  &	$	2s(^{2}S)2p^{2}	$	&	$	^{4}P_{5/2}	$	&	1154280	&	1154230	&	1154024	&	1155028	&	1153643	&	1154279	\\
6	  &	$	2s(^{2}S)2p^{2}	$	&	$	^{2}D_{3/2}	$	&	1429450	&	1429411	&	1429467	&	1436632	&	1436820	&	1429450	\\
7	  &	$	2s(^{2}S)2p^{2}	$	&	$	^{2}P_{1/2}	$	&	1502900	&	1503506	&	1503582	&	1512489	&	1513330	&	1502896	\\
8	  &	$	2s(^{2}S)2p^{2}	$	&	$	^{2}D_{5/2}	$	&	1676630	&	1670690	&	1670591	&	1674123	&	1673658	&	1676630	\\
9	  &	$	2s(^{2}S)2p^{2}	$	&	$	^{2}S_{1/2}	$	&	2029440	&	2029008	&	2029150	&	2036725	&	2037505	&	2029439	\\
10	&	$	2s(^{2}S)2p^{2}	$	&	$	^{2}P_{3/2}	$	&	2039330	&	2039964	&	2039985	&	2048688	&	2049838	&	2039329	\\
11	&	$	2p^{3}	        $	&	$	^{4}S_{3/2}	$	&		      &	2256106	&	2256094	&	2267543	&	2266564	&	2266415	\\
12	&	$	2p^{3}	        $	&	$	^{2}D_{3/2}	$	&		      &	2634142	&	2634071	&	2644684	&	2643753	&	2633355	\\
13	&	$	2p^{3}	        $	&	$	^{2}D_{5/2}	$	&	2743300	&	2732041	&	2732069	&	2744892	&	2744609	&	2743300	\\
14	&	$	2p^{3}	        $	&	$	^{2}P_{1/2}	$	&		      &	2932576	&	2932765	&	2948441	&	2949242	&	2937634	\\
15	&	$	2p^{3}	        $	&	$	^{2}P_{3/2}	$	&	3306800	&	3308788	&	3308840	&	3321726	&	3322091	&	3306801	\\

\hline \\
\end{longtable}

\newpage
\setlength{\tabcolsep}{0.5\tabcolsep}
\renewcommand{\arraystretch}{0.7}
\footnotesize 
\begin{longtable}{@{\extracolsep\fill}lllllrrrrrrrrrrl@{}}
\caption{Comparisons between the observed wavelengths (in ${\rm{\AA}}$), and the present MCDHF and RMBPT values in Kr XXXII. a - the present results; b - the NIST values~\cite{kramida2015nist}; c - Denne et al.~\cite{denne1989spectrum}; d - Martin et al.~\cite{martin19902s}; e - Myrn{\"a}s et al.~\cite{myrnas1994transitions}; f - Podpaly~\cite{Podpaly2014}; g - Kukla et al.~\cite{kukla2005extreme}. The NIST~\cite{kramida2015nist} compiled wavelengths are derived from~\cite{denne1989spectrum,martin19902s,myrnas1994transitions}. The first column ``Tag" is a label for each transition. \vskip0.5pc
\label{tab_compare_wavelengths}}
Tag & \multicolumn{4}{c}{Transitions} & RMBPT$^{a}$ &  MCDHF$^{a}$  &  NIST$^{b}$  & Observed$^{c}$ &  Observed$^{d}$ & Observed$^{e}$ &Observed$^{f,g}$       \\
\cline{2-5}\noalign{\vskip7pt}
 &   &   $i$ &  &   $j$  &   &   &   &   &   &  &         \\
\hline\\
\endfirsthead\\
\caption[]{($continued$)}
Tag & \multicolumn{4}{c}{Transitions} & RMBPT$^{a}$ &  MCDHF$^{a}$  &  NIST$^{b}$  & Observed$^{c}$ &  Observed$^{d}$ & Observed$^{e}$ &Observed$^{f,g}$       \\
\cline{2-5}\noalign{\vskip7pt}
 &   &   $i$ &  &   $j$  &   &   &   &   &   &  &         \\
\hline\\
\endhead
A &	$	2s^{2}2p	      $	&	$	^{2}P_{1/2}	$	  &	$	2s^{2}2p	      $	&	$	^{2}P_{3/2}	$	& 203.003  &		203.025	&	$203.021 ^{c}  $& $203.021 \pm 0.02  $&  $                $ &  $                 $ & $                 	   $  \\
B &	$	2s^{2}2p	      $	&	$	^{2}P_{1/2}	$	  &	$	2s(^{2}S)2p^{2}	$	&	$	^{4}P_{1/2}	$	& 143.320  &		143.352	&	$143.266	^{e} $& $    	             $&  $                $ &  $143.266 \pm 0.010$ & $                     $  \\
C &	$	2s^{2}2p	      $	&	$	^{2}P_{3/2}	$	  &	$	2s(^{2}S)2p^{2}	$	&	$	^{4}P_{5/2}	$	& 151.143  &		151.178	&	$151.121 ^{e}  $& $    	             $&  $                $ &  $151.121 \pm 0.025$ & $                     $  \\
D &	$	2s^{2}2p	      $	&	$	^{2}P_{1/2}	$	  &	$	2s(^{2}S)2p^{2}	$	&	$	^{2}D_{3/2}	$	& 69.9589  &		69.9561	&	$69.957  ^{c}  $& $69.957 \pm 0.02	 $&  $69.84 \pm 0.05  $ &  $                 $ & $69.945 \pm 0.024^{f} $  \\
E &	$	2s^{2}2p	      $	&	$	^{2}P_{1/2}	$	  &	$	2s(^{2}S)2p^{2}	$	&	$	^{2}P_{1/2}	$	& 66.5112  &		66.5078	&	$66.538  ^{c}  $& $66.538 \pm 0.025  $&  $66.49 \pm 0.05  $ &  $                 $ & $66.506 \pm 0.025^{f} $	 \\
F &	$	2s^{2}2p	      $	&	$	^{2}P_{3/2}	$	  &	$	2s(^{2}S)2p^{2}	$	&	$	^{2}D_{5/2}	$	& 84.8834  &		84.8868	&	$84.454  ^{e}  $& $84.94 \pm 0.10    $&  $84.89 \pm 0.05  $ &  $84.454 \pm 0.025 $ & $ 84.67 \pm 0.19^{g}  $  \\
G &	$	2s^{2}2p	      $	&	$	^{2}P_{3/2}	$	  &	$	2s(^{2}S)2p^{2}	$	&	$	^{2}S_{1/2}	$	& 65.0870  &		65.0788	&	$65.067  ^{e}  $& $                  $&  $65.00 \pm 0.2   $ &  $65.067 \pm 0.025 $ & $                  	 $  \\
H &	$	2s^{2}2p	      $	&	$	^{2}P_{3/2}	$		&	$	2s(^{2}S)2p^{2}	$	&	$	^{2}P_{3/2}	$	& 64.6262  &		64.6231	&	$64.651 	^{e} $& $64.65 \pm 0.10    $&  $64.59 \pm 0.2   $ &  $64.651 \pm 0.010 $ & $                     $  \\
I &	$	2s(^{2}S)2p^{2} $	&	$	^{2}D_{3/2}	$	  &	$	2p^{3}        	$	&	$	^{2}D_{3/2}	$	& 83.0060  &		83.0149	&	$          	   $& $                  $&  $                $ &  $                 $ & $82.96 \pm 0.25^{g}   $  \\
J &	$	2s(^{2}S)2p^{2}	$	&	$	^{2}P_{1/2}	$		&	$	2p^{3}	        $	&	$	^{2}D_{3/2}	$	& 88.4458  &		88.4573 &	$        	     $& $    	             $&  $                $ &  $                 $ & $88.23 \pm 0.25^{g}   $  \\
K &	$	2s(^{2}S)2p^{2}	$	&	$	^{2}D_{5/2}	$		&	$	2p^{3}	        $	&	$	^{2}D_{5/2}	$	& 94.2195  &		94.2082	&	$93.75   ^{d}  $& $    	             $&  $93.75 \pm 0.2   $ &  $                 $ & $94.11 \pm 0.25^{g}   $ 	\\
L &	$	2s(^{2}S)2p^{2}	$	&	$	^{2}P_{3/2}	$		&	$	2p^{3}	        $	&	$	^{2}P_{3/2}	$	& 78.8131  &		78.8112	&	$78.90   ^{d}  $& $    	             $&  $78.90 \pm 0.2   $ &  $                 $ & $                     $    \\

\hline \\
\end{longtable}

\normalsize

\clearpage

\newpage

\TableExplanation

\renewcommand{\arraystretch}{1.0}
\bigskip
\section*{Table ~\ref{tab_energy}.\label{tabexp_energy} Energies (in cm$^{-1}$) relative to the ground state and lifetimes ($\tau$ , in s) of RMBPT and MCDHF in Kr XXXII.}
\begin{tabular*}{1.00\textwidth}{@{}@{\extracolsep{\fill}}lp{5.5in}@{}}
Key             &A number assigned to each level, which will be used in the following tables. \\
Conf.             &The configuration. \\
$LSJ$             &The $LSJ$-coupled labels. \\
$E_{\rm RMBPT}$             &The RMBPT excitation energy, in cm$^{-1}$. \\
$E_{\rm MCDHF}$             &The MCDHF excitation energy, in cm$^{-1}$. \\
$\tau_{\rm RMBPT}$          &The RMBPT lifetime, in s. \\
$\tau_{\rm MCDHF}$          &The MCDHF lifetime, in s. \\
$LSJ$-Mixing Coefficients            &The $LS$-coupling mixing coefficients. \\
\end{tabular*}

\hangafter 0
\hangindent 1.5em
\noindent
$Note:$ The $jj$-coupled labels and mixing coefficients are given in the supplements. Table~\ref{tab_energy} is available in its entirety on the $ADNDT$ website.\\


\renewcommand{\arraystretch}{1.0}
\bigskip
\section*{Table ~\ref{tab_tr}.\label{tabexp_tr} Wavelengths ($\lambda$, in \AA), radiative rates ($A$, in s$^{-1}$), oscillator strengths ($f$, dimensionless) and line strengths ($S$, in a.u.) for transitions with a radiative branching ratio larger than 0.1\% among the 513 levels listed in Table~\ref{tab_energy}.}
\begin{tabular*}{0.95\textwidth}{@{}@{\extracolsep{\fill}}lp{5.5in}@{}}
$i$         &The lower level of a transition. \\
$j$         &The upper level of a transition. \\
$\lambda_{\rm{RMBPT}}$         &The RMBPT wavelength, in \AA. \\
$\lambda _{\rm{MCDHF}}$         &The MCDHF wavelength, in \AA. \\
Type                   &Transition type (E1, M1, E2, M2, E3).\\
$A_{\rm{RMBPT}}$         &The RMBPT transition rate, in s$^{-1}$.\\
$A^l_{\rm{MCDHF}}$         &The MCDHF transition rate in the length gauge, in s$^{-1}$.\\
$A^v_{\rm{MCDHF}}$         &The MCDHF transition rate in the velocity gauge, in s$^{-1}$.\\
BRs   &Branching ratios of the transition. \\
\end{tabular*}

\hangafter 0
\hangindent 1.5em
\noindent
$Note:$ Only the transitions among the $n = 2$ levels are shown here. Table~\ref{tab_tr} is available in its entirety on the $ADNDT$ website.\\

\renewcommand{\arraystretch}{1.0}
\bigskip
\section*{Table ~\ref{tab_collisionstrengths}.\label{tabexp_collisionstrengths} Transition energies $\Delta E$ (cm$^{-1}$), total transition rates $A$(s$^{-1}$) of E1, M1, E2, M2, E3 transitions and effective collision strengths $\Upsilon$ at ${T}_{e} = 5.12\times10^{5}, 1.02\times10^{6}, 2.05 \times10^{6}, 5.12 \times10^{6}, 1.02 \times10^{7}, \cdots , 2.05 \times10^{9}$ K.}
\begin{tabular*}{0.95\textwidth}{@{}@{\extracolsep{\fill}}lp{5.5in}@{}}
$i$         &The lower level of a transition. \\
$g_{i}$     &The statistical weights of lower levels.\\
$j$         &The upper level of a transition. \\
$g_{j}$     &The statistical weights of upper levels.\\
$\Delta E$         &The transition energy, in cm$^{-1}$. \\
$A_{ji}$         &The transition rate, in s$^{-1}$. \\
$\Upsilon / {T}_{e}$   &The effective collision strength $\Upsilon$ listed at electron temperatures $T_{e}$ in K.\\
\end{tabular*}

\hangafter 0
\hangindent 1.5em
\noindent
$Note:$ Only the arising from the ground state are given here. Table~\ref{tab_collisionstrengths} is available in its entirety on the $ADNDT$ website

\renewcommand{\arraystretch}{1.0}
\bigskip
\section*{Table ~\ref{tab_CRM}.\label{tabexp_CRM} Strong lines (with relative intensities to the strongest line larger than 0.1\%) from our CRM simulation at electron temperature of 3.5 keV. The intensity here is the relative intensity to the strongest transition at $\rho = 10^{12}, 10^{13}, \cdots, 10^{18}\ \rm{cm}^{-3}$.}
\begin{tabular*}{0.95\textwidth}{@{}@{\extracolsep{\fill}}lp{5.5in}@{}}
$i-j$         &The lower and upper levels of a transition. \\
$\lambda _{\rm{RMBPT}}$         &The RMBPT wavelength, in \AA. \\
$\lambda _{\rm{MCDHF}}$         &The MCDHF wavelength, in \AA. \\
Relative Intensity/$\rho\ (\rm{cm}^{-3}$)     &The relative intensity to the strongest transition in different electron densities $\rho\ (\rm{cm}^{-3}$) at electron temperature of 3.5 keV.\\
Type                   &Transition type (E1, M1, E2, M2, E3).\\
\end{tabular*}
                \\
                \\

\renewcommand{\arraystretch}{1.0}
\bigskip

\datatables 


\setlength{\LTleft}{0pt} \setlength{\LTright}{0pt}

\setlength{\tabcolsep}{0.5\tabcolsep}
\renewcommand{\arraystretch}{1.0}
\footnotesize 

\begin{longtable}{@{\extracolsep\fill}lllrrrrrrrcc@{}}
\caption{Energies (in cm$^{-1}$) relative to the ground state and lifetimes ($\tau$ , in s) of RMBPT and MCDHF in Kr XXXII. The last column denotes the $LSJ$ mixing coefficients from the present RMBPT calculation. See page\
\pageref{tabexp_energy} for Explanation of Tables
\label{tab_energy} }
	Key  & Conf. & $LSJ$ & $E\rm_{RMBPT}$ &  $E\rm_{MCDHF}$   &  $\tau\rm_{RMBPT}$  & $\tau\rm_{MCDHF}$ &$LSJ$  -&Mixing Coefficients &  \\
\hline\\
\endfirsthead\\
\caption[]{($continued$)}
	Key  & Conf. & $LSJ$ & $E\rm_{RMBPT}$ &  $E\rm_{MCDHF}$   &  $\tau\rm_{RMBPT}$  & $\tau\rm_{MCDHF}$ &$LSJ$  -&Mixing Coefficients &  \\
\hline\\
\endhead																																											1&    $ 2s^{2}2p         $  & $ ^{2}P_{  1/2}  $ & 0        &0         &    &     &    $ -9.94\rm{E}-01(   1)$&$                 $&$                $\\
    2&    $ 2s^{2}2p         $  & $ ^{2}P_{  3/2}  $ & 492604   &492551    &$9.404\rm{E}-07 $&$ 9.401\rm{E}-07 $&    $ -9.90\rm{E}-01(   2)$&$                 $&$                $\\
    3&    $ 2s(^{2}S)2p^{2}  $  & $ ^{4}P_{  1/2}  $ & 697742   &697583    &$8.026\rm{E}-10 $&$ 8.038\rm{E}-10 $&    $  9.06\rm{E}-01(   3)$&$                 $&$                $\\
    4&    $ 2s(^{2}S)2p^{2}  $  & $ ^{4}P_{  3/2}  $ & 1013778  &1013494   &$8.921\rm{E}-09 $&$ 8.942\rm{E}-09 $&    $  9.85\rm{E}-01(   4)$&$                 $&$                $\\
    5&    $ 2s(^{2}S)2p^{2}  $  & $ ^{4}P_{  5/2}  $ & 1154230  &1154025   &$1.078\rm{E}-09 $&$ 1.080\rm{E}-09 $&    $ -8.36\rm{E}-01(   5)$&$  5.48\rm{E}-01(   8) $&$-6.33\rm{E}-03( 103) $\\
    6&    $ 2s(^{2}S)2p^{2}  $  & $ ^{2}D_{  3/2}  $ & 1429411  &1429468   &$2.077\rm{E}-11 $&$ 2.076\rm{E}-11 $&    $  9.04\rm{E}-01(   6)$&$                 $&$                $\\
    7&    $ 2s(^{2}S)2p^{2}  $  & $ ^{2}P_{  1/2}  $ & 1503507  &1503583   &$1.004\rm{E}-11 $&$ 1.003\rm{E}-11 $&    $ -8.54\rm{E}-01(   7)$&$                 $&$                $\\
    8&    $ 2s(^{2}S)2p^{2}  $  & $ ^{2}D_{  5/2}  $ & 1670691  &1670591   &$7.127\rm{E}-11 $&$ 7.119\rm{E}-11 $&    $  8.36\rm{E}-01(   8)$&$  5.48\rm{E}-01(   5) $&$-9.92\rm{E}-03( 103) $\\
\end{longtable}

\setlength{\tabcolsep}{0.5\tabcolsep}
\renewcommand{\arraystretch}{1.0}
\setlength{\tabcolsep}{11pt}
\setlength{\abovecaptionskip}{0pt}%
\setlength{\belowcaptionskip}{10pt}%
\footnotesize 

\begin{center}
\begin{longtable}{@{\extracolsep\fill}llccccccrrrrrrr@{}}
\caption{Wavelengths ($\lambda$, in \AA), radiative rates ($A$, in s$^{-1}$), oscillator strengths ($f$, dimensionless) and line strengths ($S$, in a.u.) for transitions with a radiative branching ratio larger than 0.1\% among the 513 levels listed in Table~\ref{tab_energy}. See page\
\pageref{tabexp_tr} for Explanation of Tables
\label{tab_tr}}
$i$  & ~~$j$ & ~$\lambda_{\rm{RMBPT}}$ &  $\lambda_{\rm{MCDHF}}$  & Type &  $A_{\rm{MBPT}}$  &  $A^l_{\rm{MCDHF}}$ & $A^v_{\rm{MCDHF}}$   &  BRs~~~~\\
\hline\\
\endfirsthead\\
\caption[]{($continued$)}
$i$  & ~~$j$ & ~$\lambda_{\rm{RMBPT}}$ &  $\lambda_{\rm{MCDHF}}$  & Type &  $A_{\rm{MBPT}}$  &  $A^l_{\rm{MCDHF}}$ & $A^v_{\rm{MCDHF}}$   &  BRs~~~~\\
\hline\\
\endhead
1	&	2	&	$	2.03003\rm{E}+02	$	&	$	2.03025\rm{E}+02	$	&	M1	&	$	1.063\rm{E}+06	$	&	$	1.063\rm{E}+06	$	&	$		$	&	$	1.00\rm{E}+00	$	\\
1	&	3	&	$	1.43320\rm{E}+02	$	&	$	1.43352\rm{E}+02	$	&	E1	&	$	1.240\rm{E}+09	$	&	$	1.238\rm{E}+09	$	&	$	1.293\rm{E}+09	$	&	$	9.95\rm{E}-01	$	\\
2	&	3	&	$	4.87476\rm{E}+02	$	&	$	4.87730\rm{E}+02	$	&	E1	&	$	5.910\rm{E}+06	$	&	$	5.902\rm{E}+06	$	&	$	6.473\rm{E}+06	$	&	$	4.74\rm{E}-03	$	\\
1	&	4	&	$	9.86409\rm{E}+01	$	&	$	9.86686\rm{E}+01	$	&	E1	&	$	4.871\rm{E}+07	$	&	$	4.856\rm{E}+07	$	&	$	4.879\rm{E}+07	$	&	$	4.35\rm{E}-01	$	\\
2	&	4	&	$	1.91874\rm{E}+02	$	&	$	1.91960\rm{E}+02	$	&	E1	&	$	6.281\rm{E}+07	$	&	$	6.270\rm{E}+07	$	&	$	6.927\rm{E}+07	$	&	$	5.60\rm{E}-01	$	\\
3	&	4	&	$	3.16419\rm{E}+02	$	&	$	3.16545\rm{E}+02	$	&	M1	&	$	5.752\rm{E}+05	$	&	$	5.747\rm{E}+05	$	&	$		$	&	$	5.13\rm{E}-03	$	\\
2	&	5	&	$	1.51143\rm{E}+02	$	&	$	1.51178\rm{E}+02	$	&	E1	&	$	9.276\rm{E}+08	$	&	$	9.263\rm{E}+08	$	&	$	9.792\rm{E}+08	$	&	$	1.00\rm{E}+00	$	\\
1	&	6	&	$	6.99589\rm{E}+01	$	&	$	6.99561\rm{E}+01	$	&	E1	&	$	4.745\rm{E}+10	$	&	$	4.747\rm{E}+10	$	&	$	4.759\rm{E}+10	$	&	$	9.85\rm{E}-01	$	\\
\hline
\end{longtable}
\end{center}
$Note:$ Only the transitions among the $n = 2$ levels are shown here. Table~\ref{tab_tr} is available in its entirety on the $ADNDT$ website

\setlength{\tabcolsep}{0.5\tabcolsep}
\renewcommand{\arraystretch}{1.0}
\footnotesize 

\begin{longtable}{@{\extracolsep\fill}rrccccccccccccccl@{}}
\caption{Transition energies $\Delta E$ (cm$^{-1}$), total transition rates $A$(s$^{-1}$) of E1, M1, E2, M2, E3 transitions and effective collision strengths $\textit{$\Upsilon$}$ at $T_{e} = 5.12\times10^{5}, 1.02\times10^{6}, 2.05 \times10^{6}, 5.12 \times10^{6}, 1.02 \times10^{7}, \cdots , 2.05 \times10^{9}$ K. See page\
\pageref{tabexp_collisionstrengths} for Explanation of Tables
\label{tab_collisionstrengths} }
$i$ &$g_{i}$  & $j$ &$g_{j}$  & $\Delta E$   & $A_{ji}$   & \multicolumn{7}{c}{$\Upsilon$ / $T_{e}$}   \\
\cline{7-13}\noalign{\vskip7pt}
	&  &   &   &   &   & 5.12$\times10^{5}$  & 1.02$\times10^{6}$ & 2.05$\times10^{6}$  & 5.12$\times10^{6}$ & 1.02$\times10^{7}$ &$\cdots$  & 2.05$\times10^{9}$  \\
\hline\\
\endfirsthead\\
\caption[]{($continued$)}
$i$ &$g_{i}$  & $j$ &$g_{j}$  & $\Delta E$   & $A_{ji}$   & \multicolumn{7}{c}{$\Upsilon$ / $T_{e}$}   \\
\cline{7-13}\noalign{\vskip7pt}
	&  &   &   &   &   & 5.12$\times10^{5}$  & 1.02$\times10^{6}$ & 2.05$\times10^{6}$  & 5.12$\times10^{6}$ & 1.02$\times10^{7}$ &$\cdots$  & 2.05$\times10^{9}$  \\
\hline\\
\endhead																																											
1	&	2	&	2	&	4	&	$	4.92604\rm{E}+05	$	&	$	1.06\rm{E}+06	$	&	$	3.33\rm{E}-02	$	&	$	3.06\rm{E}-02	$	&	$	2.74\rm{E}-02	$	&	$	2.61\rm{E}-02	$	&	$	2.41\rm{E}-02	$	&	$	\cdots	$	&	$	1.47\rm{E}-02	$	\\
1	&	2	&	3	&	2	&	$	6.97742\rm{E}+05	$	&	$	1.24\rm{E}+09	$	&	$	2.47\rm{E}-02	$	&	$	2.38\rm{E}-02	$	&	$	2.37\rm{E}-02	$	&	$	2.54\rm{E}-02	$	&	$	2.71\rm{E}-02	$	&	$	\cdots	$	&	$	1.01\rm{E}-01	$	\\
1	&	2	&	4	&	4	&	$	1.01378\rm{E}+06	$	&	$	4.87\rm{E}+07	$	&	$	6.90\rm{E}-03	$	&	$	5.46\rm{E}-03	$	&	$	4.93\rm{E}-03	$	&	$	5.46\rm{E}-03	$	&	$	5.31\rm{E}-03	$	&	$	\cdots	$	&	$	2.46\rm{E}-03	$	\\
1	&	2	&	5	&	6	&	$	1.15423\rm{E}+06	$	&	$	2.88\rm{E}+02	$	&	$	6.41\rm{E}-03	$	&	$	5.28\rm{E}-03	$	&	$	5.18\rm{E}-03	$	&	$	6.25\rm{E}-03	$	&	$	6.13\rm{E}-03	$	&	$	\cdots	$	&	$	1.33\rm{E}-04	$	\\
1	&	2	&	6	&	4	&	$	1.42941\rm{E}+06	$	&	$	4.75\rm{E}+10	$	&	$	1.49\rm{E}-01	$	&	$	1.51\rm{E}-01	$	&	$	1.55\rm{E}-01	$	&	$	1.66\rm{E}-01	$	&	$	1.78\rm{E}-01	$	&	$	\cdots	$	&	$	7.92\rm{E}-01	$	\\
1	&	2	&	7	&	2	&	$	1.50351\rm{E}+06	$	&	$	9.75\rm{E}+10	$	&	$	1.30\rm{E}-01	$	&	$	1.31\rm{E}-01	$	&	$	1.34\rm{E}-01	$	&	$	1.43\rm{E}-01	$	&	$	1.54\rm{E}-01	$	&	$	\cdots	$	&	$	6.95\rm{E}-01	$	\\
1	&	2	&	8	&	6	&	$	1.67069\rm{E}+06	$	&	$	2.85\rm{E}+02	$	&	$	8.38\rm{E}-04	$	&	$	7.53\rm{E}-04	$	&	$	8.34\rm{E}-04	$	&	$	1.06\rm{E}-03	$	&	$	1.02\rm{E}-03	$	&	$	\cdots	$	&	$	2.30\rm{E}-05	$	\\
1	&	2	&	9	&	2	&	$	2.02901\rm{E}+06	$	&	$	7.69\rm{E}+08	$	&	$	4.83\rm{E}-04	$	&	$	4.79\rm{E}-04	$	&	$	5.27\rm{E}-04	$	&	$	6.28\rm{E}-04	$	&	$	6.61\rm{E}-04	$	&	$	\cdots	$	&	$	2.03\rm{E}-03	$	\\
\hline
\end{longtable}
Only transitions arising from the ground state are given here. Table~\ref{tab_collisionstrengths} is available in its entirety on the $ADNDT$ website

\setlength{\tabcolsep}{0.5\tabcolsep}
\renewcommand{\arraystretch}{1.0}
\setlength{\tabcolsep}{7pt}
\footnotesize 

\begin{longtable}{@{\extracolsep\fill}llccccccccccccl@{}}
\caption{Strong lines (with relative intensities to the strongest line larger than 0.1\%) from our CRM simulation at electron temperature of 3.5 keV. The intensity here is the relative intensity to the strongest transition at $\rho = 10^{12}, 10^{13}, \cdots, 10^{18}\ \rm{cm}^{-3}$. See page\
\pageref{tabexp_CRM} for Explanation of Tables
\label{tab_CRM}}
$i - j$  &  $\lambda_{\rm{RMBPT}}$   &  $\lambda_{\rm{MCDHF}}$    & \multicolumn{7}{c}{Relative Intensity/$\rho(\rm{cm^{-3}}$)} & TP  \\
\cline{4-10}\noalign{\vskip7pt}
	&  &   & 10$^{12}$ & 10$^{13}$ & 10$^{14}$ & 10$^{15}$ & 10$^{16}$ & 10$^{17}$ & 10$^{18}$&   \\
\hline\\
\endfirsthead\\
\caption[]{($continued$)}
$i - j$  &  $\lambda_{\rm{RMBPT}}$   &  $\lambda_{\rm{MCDHF}}$    & \multicolumn{7}{c}{Relative Intensity/$\rho(\rm{cm^{-3}}$)} & TP  \\
\cline{4-10}\noalign{\vskip7pt}
	&  &   & 10$^{12}$ & 10$^{13}$ & 10$^{14}$ & 10$^{15}$ & 10$^{16}$ & 10$^{17}$ & 10$^{18}$&   \\
\hline\\
\endhead	
$1-316	        $&3.80484 &  	3.80482   &  $	1.17\rm{E}-02$&  $	1.17\rm{E}-02  $&$ 	1.17\rm{E}-02 $&$  	1.17\rm{E}-02 $&$	1.13\rm{E}-02 $&$	1.06\rm{E}-02$&$ 	1.07\rm{E}-02 $&  	E1 \\
$1-315	        $&3.80533 &  	3.80531   &  $	9.34\rm{E}-03$&  $	9.34\rm{E}-03  $&$ 	9.33\rm{E}-03 $&$  	9.29\rm{E}-03 $&$	8.98\rm{E}-03 $&$	8.37\rm{E}-03$&$ 	8.26\rm{E}-03 $&  	E1 \\
$1-307	        $&3.81082 &  	3.81084   &  $	5.74\rm{E}-03$&  $	5.74\rm{E}-03  $&$ 	5.74\rm{E}-03 $&$  	5.72\rm{E}-03 $&$	5.58\rm{E}-03 $&$	5.32\rm{E}-03$&$ 	5.31\rm{E}-03 $&  	E1 \\
$1-304	        $&3.81744 &  	3.81744   &  $	3.70\rm{E}-03$&  $	3.70\rm{E}-03  $&$ 	3.70\rm{E}-03 $&$  	3.68\rm{E}-03 $&$	3.56\rm{E}-03 $&$	3.36\rm{E}-03$&$ 	3.66\rm{E}-03 $&  	E1 \\
$1-302	        $&3.82250 &  	3.82251   &  $	2.15\rm{E}-03$&  $	2.15\rm{E}-03  $&$ 	2.15\rm{E}-03 $&$  	2.14\rm{E}-03 $&$	2.07\rm{E}-03 $&$	1.94\rm{E}-03$&$ 	1.99\rm{E}-03 $&  	E1 \\
$1-290	        $&3.88678 &  	3.88682   &  $	4.67\rm{E}-02$&  $	4.67\rm{E}-02  $&$ 	4.67\rm{E}-02 $&$  	4.68\rm{E}-02 $&$	4.71\rm{E}-02 $&$	4.78\rm{E}-02$&$ 	4.78\rm{E}-02 $&  	E1 \\
$3-318	        $&3.90276 &  	3.90277   &  $	1.17\rm{E}-03$&  $	1.17\rm{E}-03  $&$ 	1.17\rm{E}-03 $&$  	1.17\rm{E}-03 $&$	1.19\rm{E}-03 $&$	1.28\rm{E}-03$&$ 	2.12\rm{E}-03 $&  	E1 \\
$3-306	        $&3.91616 &  	3.91618   &  $	1.53\rm{E}-03$&  $	1.53\rm{E}-03  $&$ 	1.53\rm{E}-03 $&$  	1.53\rm{E}-03 $&$	1.48\rm{E}-03 $&$	1.41\rm{E}-03$&$ 	1.63\rm{E}-03 $&  	E1 \\
\hline
\end{longtable}

\end{document}